\documentclass[aps,twocolumn,noshowkeys,superscriptaddress,noshowpacs]{revtex4}

\usepackage{dcolumn}
\usepackage{amsmath}
\usepackage{bm}
\usepackage{nicefrac}
\usepackage{graphicx}
\usepackage{siunitx}
\usepackage{epsfig}
\usepackage{epstopdf}
\usepackage{xcolor}

\newcommand{\spinel}{MgAl$_\text{2}$O$_\text{4}$}

\newcommand{\nfo}{NiFe$_\text{2}$O$_\text{4}$}
\newcommand{\Ar}{Ar\,:\,O$_2$ ratio~}
\newcommand{\Zntwo}{Zn$^{2+}$}
\newcommand{\Nitwo}{Ni$^{2+}$}
\newcommand{\Fetwo}{Fe$^{2+}$}
\newcommand{\FeOhtwo}{Fe$^{2+}_{\text{Oh}}$}
\newcommand{\NiOhtwo}{Ni$^{2+}_{\text{Oh}}$}
\newcommand{\NiTdtwo}{Ni$^{2+}_{\text{Td}}$}
\newcommand{\Fethree}{Fe$^{3+}$}

\begin{document}

\title{Control of site occupancy by variation of the Zn and Al content in NiZnAl ferrite epitaxial films with low magnetic damping}
\date{\today}

\author{Julia Lumetzberger}
\email{julia.lumetzberger@jku.at; Phone: +43-732-2468-9651; FAX: -9696}
\affiliation{Institute for Semiconductor and Solid State Physics, Johannes Kepler University Linz, Altenberger Straße 69, 4040 Linz, Austria}
\author{Verena Ney}
\affiliation{Institute for Semiconductor and Solid State Physics, Johannes Kepler University Linz, Altenberger Straße 69, 4040 Linz, Austria}
\author{Anna Zhakarova}
\affiliation{Swiss Light Source (SLS), Paul Scherrer Institut, 5232 Villigen PSI, Switzerland}
\author{Daniel Primetzhofer}
\affiliation{Department of Physics and Astronomy, \si{\angstrom}ngstr{\"o}m Laboratory, Uppsala University, Box 516, SE-751 20 Uppsala, Sweden}
\author{Kilian Lenz}
\affiliation{Helmholtz-Zentrum Dresden -- Rossendorf, Institute of Ion Beam Physics and Materials Research, Bautzner Landstraße 400, 01328 Dresden, Germany}
\author{Andreas Ney}  
\affiliation{Institute for Semiconductor and Solid State Physics, Johannes Kepler University Linz, Altenberger Straße 69, 4040 Linz, Austria}

\begin{abstract}
The structural and magnetic properties of Zn/Al doped nickel ferrite thin films can be adjusted by changing the Zn and Al content. The films are epitaxially grown by reactive magnetron sputtering using a triple cluster system to sputter simultaneously from three different targets. Upon the variation of the Zn content the films remain fully strained with similar structural properties, while the magnetic properties are strongly affected. The saturation magnetization and coercivity as well as resonance position and linewidth from ferromagnetic resonance (FMR) measurements are altered depending on the Zn content in the material. The reason for these changes can be elucidated by investigation of the x-ray magnetic circular dichroism spectra to gain site and valence specific information with elemental specificity. Additionally, from a detailed investigation by broadband FMR a minimum in g-factor and linewidth could be found as a function of film thickness. Furthermore, the results from a variation of the Al content using the same set of measurement techniques is given. Other than for Zn, the variation of Al affects the strain and even more pronounced changes to the magnetic properties are apparent.
\end{abstract}

\maketitle

\section*{I. Introduction}
Zn/Al doped nickel ferrites (NiZAF) came into the focus of recent research because of their promising characteristics for application in spintronics, e.g. spin pumping \cite{WA01, TBB02}. The material is insulating and ferromagnetic at room temperature with low magnetic damping. Therefore, it can be considered as a good candidate \cite{E17, L20, D21} among few other materials, e.g. MAFO \cite{W20, R19}, all-perovskite oxides \cite{E16} and Fe alloyed with V and Al \cite{S20} as replacement, for the well known magnetic insulator yttrium iron garnet (YIG) \cite{OKK14, CLZ14}. YIG exhibits comparable magnetic properties as NiZAF, however, it has the complex garnet structure and thin film growth requires gadolinium gallium garnet (GGG) as substrates thus limiting a wide applicability. Since the number of ferromagnetic insulators with low magnetic damping is sparse \cite{HE15}, the search for appropriate replacements for YIG such as NiZAF is still ongoing and of high interest in current spintronic research. \\
The ideal nominal stoichiometry of NiZAF was reported to be Ni$_\text{0.65}$Zn$_\text{0.35}$Al$_\text{0.8}$Fe$_\text{1.2}$O$_\text{4}$. It is based on theoretical values and experiments on the bulk material \cite{L13,D11}. NiZAF grows in the cubic spinel structure and is derived from \nfo, which grows in the inverse spinel crystal structure  [Fe$_{1}$]$^{\text{A}}$[Ni$_{1}$Fe$_{1}$]$^{\text{B}}$O$_4$, where A denotes the tetrahedral (Td) and B denotes the octahedral (Oh) site. In the ideal case Ni only occupies B sites and Fe sits in a 1:1 ratio on the A and B sites. However as previously reported \cite{E17,L20}, due to the growth on a normal spinel substrate \spinel(001) and by introducing the two additional elements Zn and Al a mixed spinel is the result. Therefore, Ni can be considered to also occupy tetrahedral sites and additionally \FeOhtwo \, was found to be introduced into the system \cite{E17, L20}. In addition to this complex interplay of several elements with different valencies, charge neutrality has to be considered as well, and small deviations can be compensated to some extent by the formation of oxygen vacancies.\\
According to theory \cite{D11}, a variation of the Zn content will procure the biggest changes to the magnetic properties since it is incorporated in the crystal structure with a valence of 2+, preferably on tetrahedral sites. Therefore, it can be supposed to mainly affect \Nitwo\, and \Fetwo. In particular, since \NiTdtwo \, and \FeOhtwo \, have the biggest negative impact on the magnetic properties, i.e., damping by unquenched orbital moment and hopping [Fe$^{2+}$ $\rightarrow$ Fe$^{3+}$ + e\text{-}], respectively.  A systematic variation of Zn is therefore expected to show how the occupancy and/or valence of the cations change and affect the structural properties, and more importantly the magnetic properties. In contrast, Al doping is mainly introduced to reduce the strain of epitaxial thin films. Even though the Al content was already varied in bulk, showing a dependence of the saturation magnetization \cite{L13}, a systematic study on thin films, which may have additional effects, is still lacking.\\
The earliest reports on NiZAF thin films grown with pulsed laser deposition (PLD) revealed a highly strained soft magnetic material with a low magnetic damping \cite{E17}. However, due to the high preparation temperatures and the volatility of Zn, the thin films showed a Zn deficiency. In addition, the exact amount of Al in these PLD grown films was not reported. The Zn deficiency could be completely removed and even turned into an excess of Zn in NiZAF thin films grown by reactive magnetron sputtering (RMS) \cite{L20}, resulting in an even higher strained material with slightly increased magnetic damping. Therefore, a systematic variation of the Zn content, and in a second step, the Al content should be performed in order to optimise the composition with regard to the desired soft magnetic properties of RMS grown thin films of NiZAF.\\
In this work the composition of NiZAF was systematically changed using RMS in a triple cluster system, deviating from the nominal reported best stoichiometry \cite{E17}. The Zn content was adapted by varying the sputter power of the ZnO target. The resulting samples were analyzed for their structural properties, actual chemical composition and static and dynamic magnetic properties to determine the optimum Zn content. The valence and site occupation of the Ni and Fe cations was studied by x-ray magnetic circular dichroism (XMCD) spectra. In combination with multiplet ligand field simulations \cite{SG10}, a comparison to previous NiZAF thin films was drawn \cite{L20}. In a next step, the thickness of films with optimized Zn content was varied and the resulting series was analyzed by broadband ferromagnetic resonance (FMR) to disentangle the various contributions to the linewidth broadening and to calculate the g-factor and Gilbert damping. In a last step, the variation of the Al content was investigated for optimized Zn content using the same set of methods as mentioned above.

\section*{II. Experimental Details}

\begin{figure*}[ht]
	\centering
	\includegraphics[width=1\textwidth]{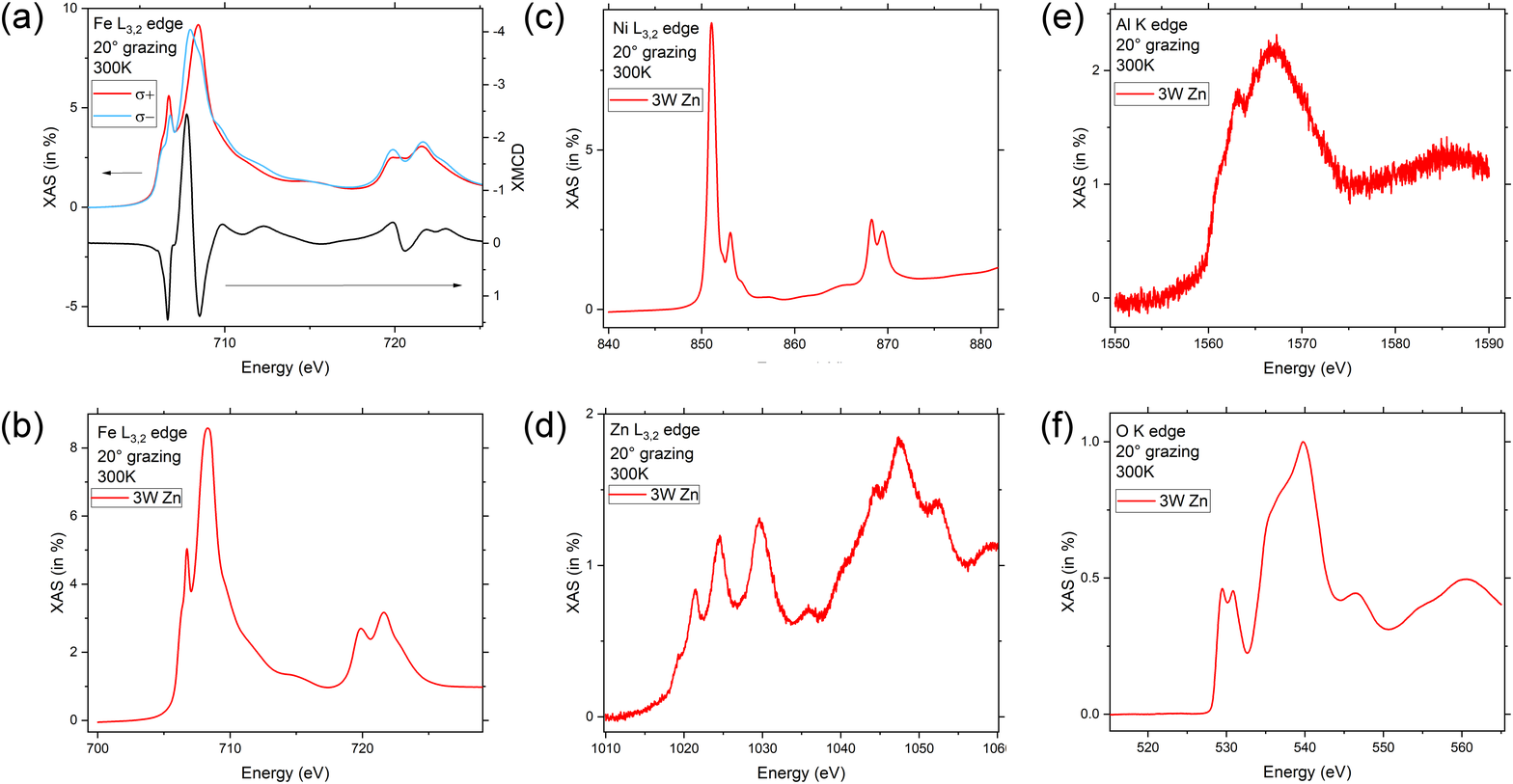}
	\vspace{-0.2cm}
	\caption{Absorption spectra recorded at \SI{300}{\kelvin} and \SI{20}{\degree} grazing from the sample with optimized growth conditions of the Zn series. (a) shows the $\sigma^+$ and $\sigma^-$ spectra and the resulting XMCD at the Fe L$_{3,2}$ edges. In (b-f)  the XAS spectra at the Fe , Ni and Zn L$_{3,2}$ edges and Al and O K edges are shown, respectively.}
	\label{figure1}
\end{figure*}

In this work NiZAF epitaxial thin films were fabricated by RMS in a triple cluster system using three different targets. The base material \nfo\, was co-deposited using ZnO and Al as additional targets. The resulting composition of NiZAF was achieved by setting different sputter powers for each target. The epitaxial thin films were grown in an ultra high vacuum (UHV) chamber with a base pressure of $4 \times 10^{-8}$\,\si{\milli\bar}. A doubleside polished single crystalline spinel [\spinel(001)] was chosen as a substrate. The optimized growth conditions are a sample temperature of \SI{650}{\degreeCelsius}, an \Ar of ($10$\,:\,$0.5$)\,standard cubic centimeter and a working pressure of $4 \times 10^{-3}$\,\si{\milli\bar}. The sputter power of the ZnO target was varied between \SIrange{3}{6}{\watt} in steps of \SI{1}{\watt},while the Al target was initially kept constant at \SI{16}{\watt} and the \nfo\, target at \SI{70}{\watt}. The parameter range was determined by recalculation from the reported best stoichiometry using the deposition rates measured by a quartz crystal micro balance. Note that, it was not feasible to create a stable plasma with a sputter power of \SI{2}{\watt} at the ZnO target, which limits the lowest achievable Zn concentration. This sample series with a nominal thickness of \SI{40}{\nano\meter} is referred to as the Zn series in the following. Additionally, on optimized preparation conditions from the Zn series the thickness (\SIrange{10}{40}{\nano\meter}) and Al content (\SIrange{15}{17}{\watt}) were varied, which are referred to as the thickness and the Al series, respectively. Note that, the sputter time for samples thinner than \SI{16}{\nano\meter} ranges from \SIrange{3}{5}{\minute}. Therefore, a thermal equilibrium after opening of the shutter and exposing the heated substrate to the plasma may not be reached during sputtering  \\
The structural characterization of the samples was done by X-ray diffraction (XRD) measurements with a \textit{Pananalytical X'Pert MRD} recording $\omega - 2 \theta$ scans and symmetric as well as asymmetric reciprocal space maps (RSM). The composition of the NiZAF thin films was determined by ion beam analysis, i.e.\ Rutherford backscattering spectrometry (RBS) using \SI{2}{\mega\electronvolt} He$^{+}$ and \SI{10}{\mega\electronvolt} $^{12}$C$^{3+}$ primary beams at the Tandem Laboratory at Uppsala University. To disentangle the elemental contributions, the spectra were analyzed using the SIMNRA software \cite{M97}. Details of the experimental set-up are described elsewhere \cite{M19}. To exclude a contamination of the material by light elements like hydrogen and carbon, electron recoil detection analysis (ERDA) using an \SI{36}{\mega\electronvolt} iodine primary beam was performed. The details of the used set-up can be found elsewhere \cite{Q17}.\\
The static magnetic properties were measured with integral superconducting quantum interference device (SQUID) magnetometry by a \textit{Quantum Design MPMS-XL5} system applying the magnetic field in the film plane (IP) as well as out-of-plane (OOP). The magnetic behavior in a field range of $\pm 5$\,\si{\tesla} from \SI{300}{\kelvin} down to \SI{2}{\kelvin} and the temperature dependence of the magnetization up to \SI{395}{\kelvin} were measured. The data were background corrected for the contribution of the diamagnetic substrate and known artifacts were carefully avoided \cite{SSA11}. In particular, the superconducting magnet was resetted to record reliable hysteresis loops at \SI{300}{\kelvin} restricting the magnetic field to below \SI{10}{\milli\tesla} \cite{B18}.\\
The dynamic magnetic properties were measured by FMR using a conventional X-band set-up (\SI{9.5}{\giga\hertz}) at \SI{300}{\kelvin}. Additionally, the samples from the thickness series were analyzed with a broadband vector network analyzer ferromagnetic resonance (VNA-FMR) set-up to determine the frequency dependence of the resonance position and FMR linewidth up to \SI{50}{\giga\hertz} and to verify the g-factor and disentangle the different contributions to the magnetic damping. Furthermore, for selected samples also the polar and azimuthal angular dependence were recorded. \\
Finally, X-ray absorption spectra (XAS) at the Ni, Fe and Zn L$_{3,2}$ edges, as well as the O and Al K edges were measured at the X-Treme beamline at the Swiss Light Source (SLS) \cite{PFS12}. The spectra were obtained in total electron yield (TEY) with \SI{20}{\degree} grazing incidence at \SI{300}{\kelvin} and a field of \SI{5}{\tesla} using circular polarized light $\sigma$. To obtain the XMCD the difference between the normalized XAS recorded with $\sigma^+$ and $\sigma^-$ light was taken. In Fig.\,\ref{figure1}(a) the $\sigma^+$ and $\sigma^-$ spectra of the best sample from the Zn series for the Fe L$_{3,2}$ edges and the resulting XMCD are shown exemplarily. The XAS is obtained by the average of the $\sigma^+$ and $\sigma^-$ spectra and shown in Fig.\,\ref{figure1}(b). The multiplet splitting at the L$_{3,2}$ edges is typical for an oxide. The XAS of the other constituents were also recorded and Fig.\,\ref{figure1} summarizes them for the Ni (c) and Zn (d) L$_{3,2}$ edges as well as the Al (e) and O (f) K-edges. All spectra were recorded with circularly polarized light, however, besides Ni no significant XMCD could be detected. For the K-edges, where the XMCD is knowingly rather small, this is understandable, in particular for the rather noisy Al K-edge. Thus, no conclusive statements about an eventual magnetic polarization of Al can be made.  On the other hand, the absence of a significant XMCD at the Zn L-edges (not shown) indicates no magnetic polarization of the Zn in the sample. The XMCD spectra at the Fe and Ni L$_{3,2}$ edges will be discussed in detail further below.

\section*{III. Structural Properties: Zn series}

\begin{figure*}[ht]
	\centering
	\includegraphics[width=1\textwidth]{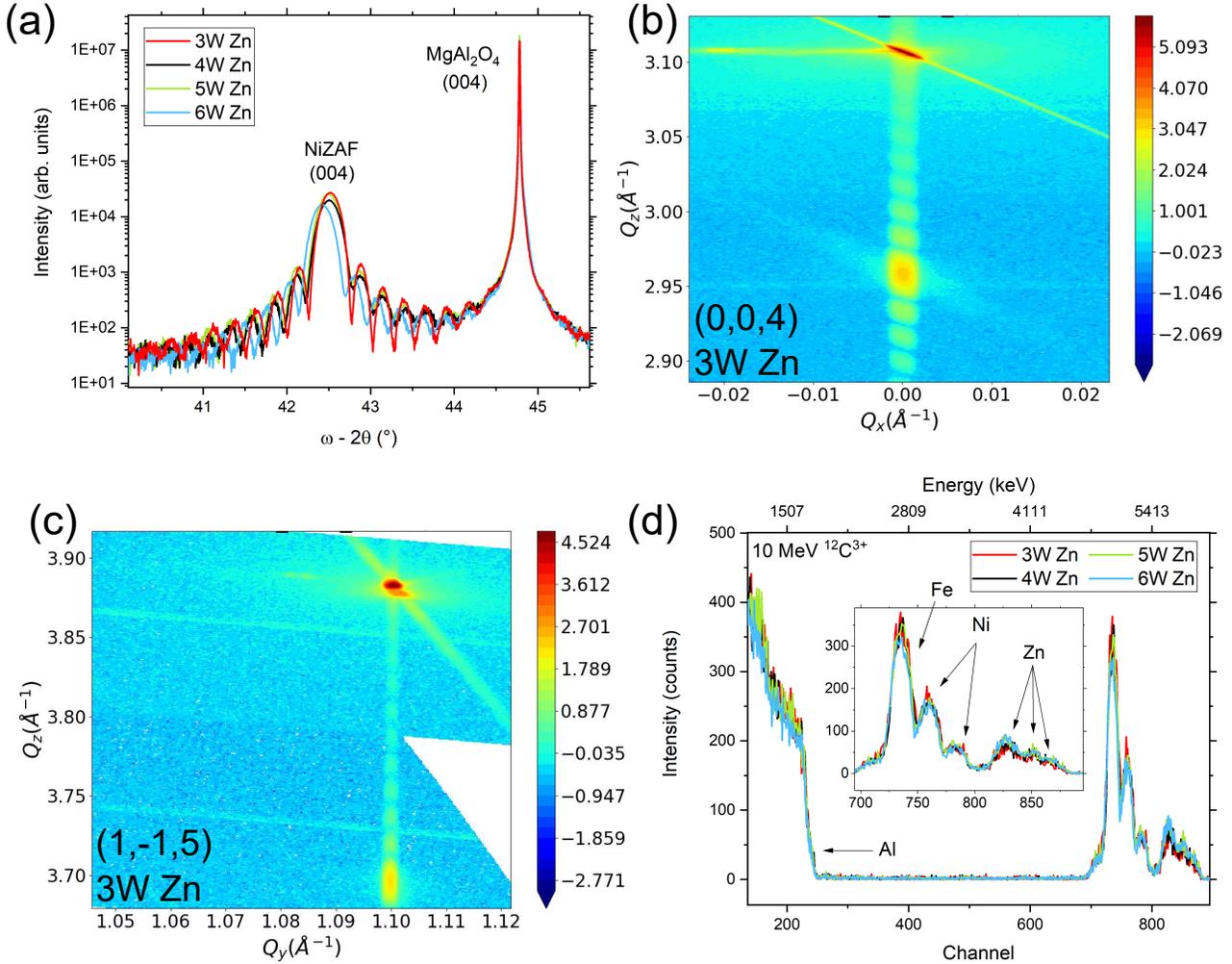}
	\vspace{-0.2cm}
	\caption{Structural analysis on the Zn series. In (a) the symmetric $\omega-2 \theta$ scans. The symmetric and asymmetric RSM of the \SI{3}{\watt} sample is shown exemplary for the Zn series in (b) and (c), respectively. In (d) the RBS spectra using a \SI{10}{\mega\electronvolt} $^{12}$C$^{3+}$ primary ion beam are shown for each sample.}
	\label{figure2}
\end{figure*} 

The samples from the Zn series were routinely analyzed by XRD and the results of the symmetric $\omega-2 \theta$ scans are shown in Fig.\,\ref{figure2}(a). The (004) reflection of the NiZAF peak has pronounced Laue oscillations, indicating a smooth surface and excellent crystal quality for each sample. The second reflection stems from the \spinel(004) substrate. The NiZAF(004) reflections for the entire series are located at an angle of $42.5$\,\si{\degree} with a full width half maximum (FWHM) of $0.25$ for the \SI{40}{\nano\meter}-thick films. This corresponds to a perpendicular lattice parameter of $a_{\perp} = (8.50 \pm 0.01)$\,\si{\angstrom}. Only the sample grown with \SI{6}{\watt} Zn shifts to slightly lower angles, but no indication for a reduced crystal quality is obvious. In Figs.\,\ref{figure2}(b) and (c) the symmetric and asymmetric reciprocal space map (RSM) of the \SI{3}{\watt} Zn sample from the Zn series are shown, respectively. Both space maps highlight the good crystal quality of the grown samples, since Laue oscillations are visibly in both cases. From the symmetric RSM along the ($004$) the peak position and calculated out-of-plane lattice parameter for NiZAF are confirmed. From the asymmetric RSM along the ($\bar{1}\bar{1}5$) plane it is apparent that the substrate and film reflection are perfectly aligned in the $Q_{\text{y}}$ axis, indicating a fully strained material in the in-plane direction. From this an in-plane lattice parameter of $a_{\parallel} = (8.08 \pm 0.01)$\,\si{\angstrom} identical to the lattice parameter of the cubic \spinel\, substrate is obtained. This results in a tetragonal distortion of  $c/a = 1.052 \pm 0.003$, which is significantly higher than reported for epitaxial NiZAF films before \cite{E17, L20, D21}. \\
To confirm the systematic variation of the Zn content by sputter power, RBS measurements were performed using a \SI{2}{\mega\electronvolt} He$^+$ ion beam (not shown) and a \SI{10}{\mega\electronvolt} $^{12}$C$^{3+}$ ion beam (see Fig.\,\ref{figure2}(d)). The advantage of using heavier ions is the better separation of Fe, Ni and Zn, which have a similar atomic mass, even revealing the splitting into the isotopes. The peaks of the energies corresponding to the different elements in the NiZAF film are indicated with arrows. From a direct comparison, differences are only evident for the Fe peak, where the \SI{6}{\watt} sample shows a reduced intensity. Any changes in Ni and Zn concentration cannot be determined by direct comparison of the spectra. The lighter element Al only shows as a tiny step at lower energies and assumptions about the composition can only be made by using SIMNRA \cite{M97} simulations to determine the percentages. The main focus is on the small variation of the Zn amount. To resolve the Zn amount an exact knowledge about the percentages of the biggest contributor oxygen would be necessary, which is not feasible using a C beam. In order to eliminate trivial dependencies such as the charge and solid-angle product of the experiment or the inelastic energy loss in the substrate material, which is needed for normalization, we are primarily discussing the respective ratios of two of the elements, which is influenced almost exclusively by the statistics of the experiment. Note, that other light species were only found at insignificant concentrations according to ERDA (not shown).\\
By increasing the sputter power of the ZnO target from \SIrange{3}{6}{\watt}, the Zn\,:\,Fe ratio increased from $0.23$ to $0.35$. The same behavior can be observed for the Zn\,:\,Ni and the Zn\,:\,Al ratio, which are increasing as well from $0.29$ to $0.61$ and from $0.37$ to $0.52$, respectively. Interestingly, an increase in Zn seems to reduce not only the Ni, but also affects the Al and Fe concentration. According to theory \cite{L13,E17,D11} Zn should only substitute for Ni. The Ni\,:\,Fe ratio on the other hand stays constant around $0.58$. Additionally, the Ni\,:\,Al and Fe\,:\,Al ratios both decrease by increasing the sputter power (not shown). Table\,\ref{table1} summarizes the results obtained for the chemical composition and the XRD analysis in the first two columns. For comparison also the findings of previous publications are provided \cite{E17,L20}. A comparison of the relative compositions shows that the results of the present sample series is in between previous results regarding the ratio of Zn to the other metals. The Zn-deficient sample grown by PLD \cite{E17} has similar ratios of Zn\,:\,Fe of $0.22$ and Zn\,:\,Ni of $0.4$ as the \SI{3}{\watt} sample. Unfortunately, the Al concentration is not reported in \cite{E17}. Ratios of Zn\,:\,Fe of $0.42$, Zn\,:\,Ni of $0.88$ and Zn\,:\,Al of $0.55$ for the previous RMS sample \cite{L20} are even higher than what was found for the \SI{6}{\watt} sample from the Zn series. From the Ni\,:\,Fe ratio the low value of $0.48$ in \cite{L20}, further highlights the abundance of Zn in the reported sample. \\
Summarizing this part, as expected, the relative composition of the elements showed that the Zn content increases with increasing sputter power. In comparison to the two previous publications on NiZAF, the ratios of Zn to the other metals of the entire Zn series are between the Zn-deficient NiZAF grown by PLD \cite{E17} and the Zn-rich NiZAF grown by RMS\cite{L20}. The structural analysis of the entire Zn series results in NiZAF films, which are even more strained compared to previous publications \cite{E17,L20, D21}. Nevertheless, all samples maintain their high crystalline quality.

\section*{IV. Magnetic Properties: Zn series}

\begin{figure}[h]
	\centering
	\includegraphics[width=0.47\textwidth]{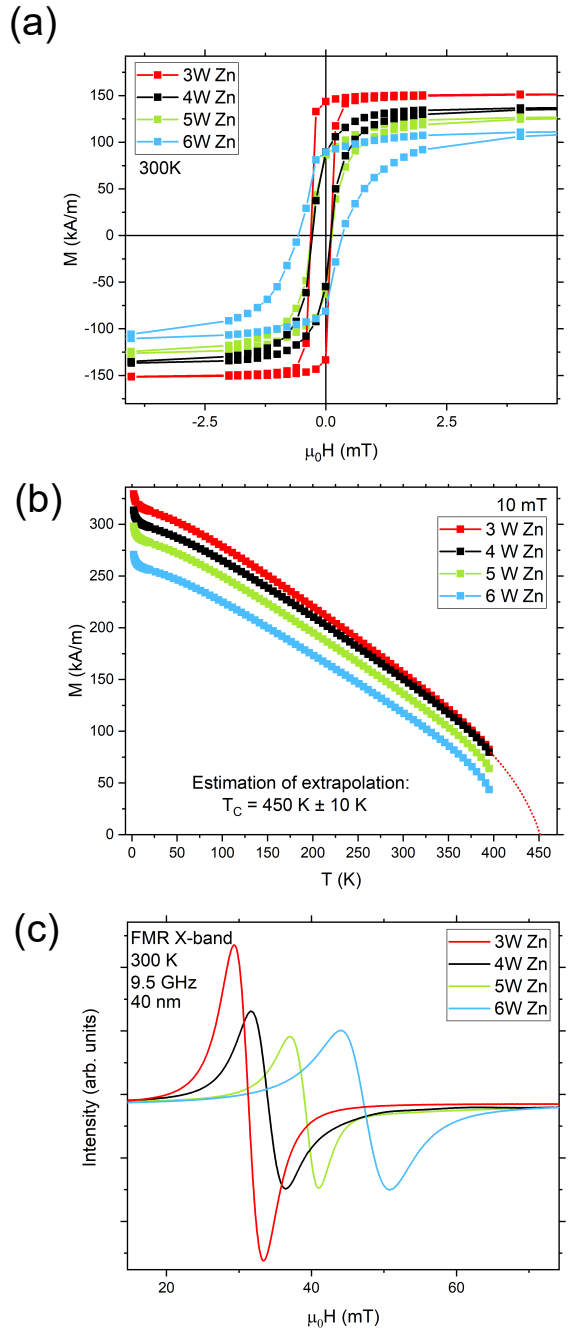}
	\vspace{-0.2cm}
	\caption{Magnetic analysis of the Zn series. In (a) the $M(H)$ curves at \SI{300}{\kelvin}, in (b) the $M(T)$ curves at \SI{10}{\milli\tesla} with an extrapolation of the Curie temperature and in (c) the X-band FMR at \SI{300}{\kelvin} and \SI{9.5}{\giga\hertz}, for each sample from the Zn series are shown.}
	\label{figure3}
\end{figure}

Having established a good crystal quality and the systematic increase in Zn content of the Zn series, static magnetic measurements were performed. In Fig.\,\ref{figure3}(a) the $M(H)$ curves at \SI{300}{\kelvin} are shown in the range of a few \si{\milli\tesla} to better visualize the hysteresis opening. A change in coercivity depending on the Zn content is apparent, where the lowest concentration present in the \SI{3}{\watt} sample leads to the smallest coercivity of lower than $\mu_0H_{\text{c}} = 0.2 $\,\si{\milli\tesla}. For a Zn amount of the \SI{6}{\watt} sample the coercivity is at least twice as high. The saturation magnetization at \SI{5}{\tesla} (not shown) is highest for the \SI{3}{\watt} sample with $M_{\text{s}} = (159 \pm 16)$\,\si{\kilo\ampere/\meter} and decreases to $M_{\text{s}} = (122 \pm 12)$\,\si{\kilo\ampere/\meter} for the \SI{6}{\watt} sample. The reduction of $M_{\text{s}}$ with increasing Zn is reasonable, since a partial substitution of Fe and Ni with the nonmagnetic Zn leads to a reduced magnetization. The $M(T)$ curves measured at \SI{10}{\milli\tesla} shown in Fig.\,\ref{figure3}(b) exhibit a similar trend for the Zn series. All Curie temperatures $T_{\text{C}}$ are well above \SI{400}{\kelvin}, however the $T_{\text{C}}$ of the \SI{6}{\watt} sample appears to be the lowest, because of the sharpest curvature of the $M(T)$ curve close to \SI{400}{\kelvin}. Measurements up to higher temperatures were not feasible since the magnetometer is limited to \SI{400}{\kelvin}. The estimated $M(T)$ behavior at higher temperatures, is illustrated exemplarily by the dotted line. From that a high $T_{\text{C}} = 450$\,\si{\kelvin} compared to previous findings for sputtered NiZAF thin films \cite{L20} was obtained for the \SI{3}{\watt} sample. The shape of the $M(T)$ curve is very similar for the entire Zn series, while the magnetization systematically decreased with increasing Zn content, which is consistent with the decrease of the saturation magnetization mentioned above. 

\begin{table*}[ht]
	\centering
	\caption{Summary of the structural and magnetic properties of the samples from the Zn series including \cite{E17,L20}. The respective measuring errors are given in the first row.}
	\begin{tabular}{c|cccc|c|cc|ccc}
		\hline \hline
		Growth & \multicolumn{4}{c|}{RBS} & XRD & \multicolumn{2}{c|}{SQUID} & \multicolumn{2}{c}{FMR} \\ 
		 Zn {(}\si{\watt}{)} &Zn:Fe & Zn:Ni& Zn:Al &Ni:Fe   & (004) reflex {(}\si{\degree}{)} & $M_{\text{s}}${(}\si{\kilo\ampere/\meter}{)} & $\mu_0H_{\text{c}}${(}\si{\milli\tesla}{)} & $\mu_0 H_{\text{res}}${(}\si{\milli\tesla}{)} & $\mu_0\Delta H_{\text{pp}}${(}\si{\milli\tesla}{)} \\ \hline \hline
		 $\pm 0.5$ {(}\si{\watt}{)} & & &  &  & $\pm 0.02$ {(}\si{\degree}{)} & $\pm 10${(}\si{\percent}{)} & $\pm 0.2${(}\si{\milli\tesla}{)} & $\pm 0.2${(}\si{\milli\tesla}{)} & $\pm 0.2${(}\si{\milli\tesla}{)} \\ 
		 3.0                    &0.23&0.39&   0.37    &0.59                               &  42.52                    & 159         & 0.2         & 31.3        & 3.9         \\
		4.0                      &0.28&0.48&   0.44  &   0.59                             & 42.50                     & 156         & 0.2         & 34.1        &  4.9           \\ 
		5.0                             &0.31&0.53&  0.48&          0.58                       & 42.50                    & 142         & 0.2         & 39.0       & 4.0             \\ 
		6.0                         &0.35&0.61&0.52&            0.58                     & 42.43                     & 122         & 0.4         &  47.3        &    6.8          \\ 
		Ref. 
		\cite{E17}                      &0.22&0.40& - &    0.55                &        43.25                &      120    &     0.2     &       84.8   &      0.8        \\ 
		Ref. \cite{L20}                  &0.42&0.88&0.55  &   0.48                           &         42.56                 &     195     & 0.2        &   57.8       &        4.4      \\ 
	\end{tabular}
	\label{table1}
\end{table*}

Furthermore, FMR measurements were performed to obtain a first indication about the size of the magnetic damping with the FMR line. A comparison of the resonances of the samples of the Zn series measured at \SI{9.5}{\giga\hertz} and \SI{300}{\kelvin} using a conventional X-band set-up is shown in Fig.\,\ref{figure3}(c). The measured resonances are fitted with a Lorentzian to obtain the resonance position ($\mu_0H_{\text{res}}$) and peak-to-peak linewidth ($\mu_0\Delta H_{\text{pp}}$). A clear increase of the resonance position with sputter power is evident, indicating reduced magnetic anisotropy with increasing Zn content. The lowest resonance position was obtained for the \SI{3}{\watt} sample, which is around $\mu_0 H_{\text{res}} = (31.3 \pm 0.2)$\,\si{\milli\tesla}. This sample also has the smallest linewidth of $\mu_0\Delta H_{\text{pp}} = (3.9 \pm 0.2)$\,\si{\milli\tesla} of the Zn series. The linewidth also increases with the Zn content reaching $\mu_0\Delta H_{\text{pp}} = (6.8 \pm 0.2)$\,\si{\milli\tesla} for the \SI{6}{\watt} sample. The results of the magnetic characterization of the Zn series together with \cite{E17,L20} are shown in Tab.\,\ref{table1} yielding the following comparison: The Zn-deficient sample from \cite{E17}, has a saturation magnetization of $M_{\text{s}} = (120 \pm 12)$\,\si{\kilo\ampere/\meter} and a small coercivity. The resonance position was found at $\mu_0 H_{\text{res}} = (84.8 \pm 0.2)$\,\si{\milli\tesla} with a very small linewidth of $\mu_0\Delta H_{\text{pp}} = (0.8 \pm 0.2)$\,\si{\milli\tesla}. The Zn-rich sample from \cite{L20} has a higher $M_{\text{s}}$ of \SI{195}{\kilo\ampere/\meter} with an equally small coercivity and reduced resonance position of $\mu_0 H_{\text{res}} = (57.8 \pm 0.2)$\,\si{\milli\tesla} with a linewidth of $\mu_0\Delta H_{\text{pp}} = (4.4 \pm 0.2)$\,\si{\milli\tesla}. Interestingly, the \SI{3}{\watt} sample with the lowest amount of Zn showed the best results from the preliminary analysis. The thin film has a saturation magnetization between the Zn-deficient \cite{E17} and Zn-rich \cite{L20} NiZAF with a comparably small coercivity. However, the resonance position is drastically shifted to $\mu_0 H_{\text{res}} = (31.3 \pm 0.2)$\,\si{\milli\tesla}. This can be correlated with the higher strain in this sample favoring higher magnetic anisotropy and thus a lower resonance field. Interestingly, the Curie temperature of \SI{450}{\kelvin} found for the \SI{3}{\watt} sample is so far the highest reported value for sputtered NiZAF thin films and matches the result from PLD grown samples in \cite{E17}. Although, the sample is even more strained than the reported Zn-deficient \cite{E17} and Zn-rich \cite{L20} samples, the linewidth remained narrow with $\mu_0\Delta H_{\text{pp}} = (3.9 \pm 0.2)$\,\si{\milli\tesla}. Obviously, the \SI{3}{\watt} sample has the most promising Zn concentration with regard to the desired soft magnetic properties. In a next step, the cation distribution of the magnetic cations in NiZAF will be analyzed for the entire Zn series.

\section*{V. Cation distribution: Zn series}

\begin{figure}[h]
	\centering
	\includegraphics[width=0.45\textwidth]{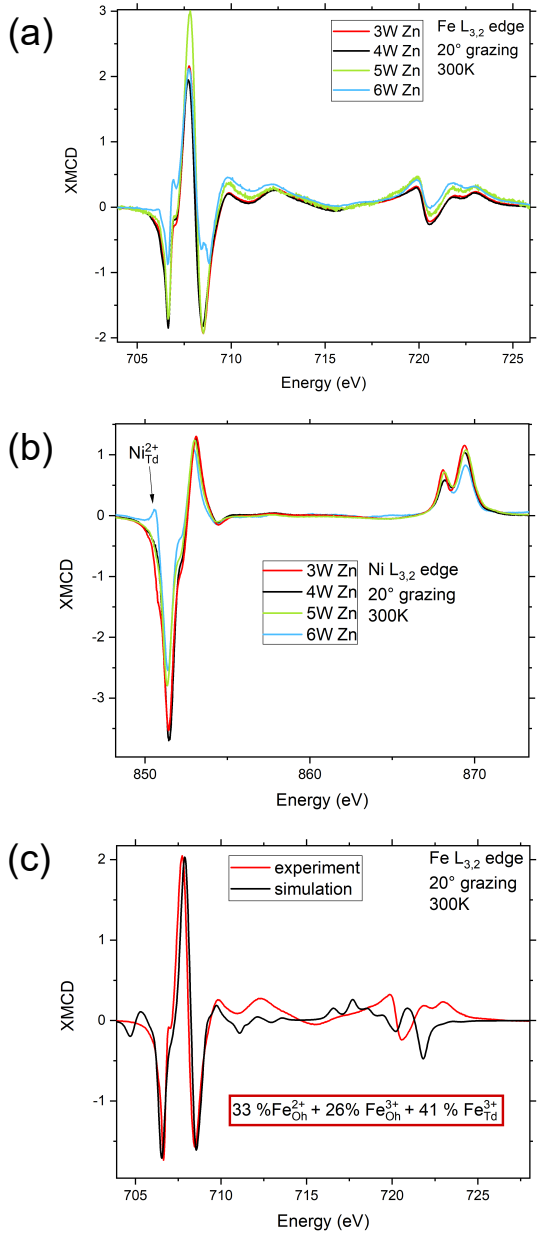}
	\vspace{-0.2cm}
	\caption{XMCD spectra of the samples from the Zn series. In (a) the XMCD spectra at \SI{300}{\kelvin} in \SI{20}{\degree} grazing at the Fe L$_{3,2}$ edges and in (b) at Ni L$_{3,2}$ edges are shown. (c) depicts a comparison of the multiplet ligand field simulation using CTM4XAS \cite{SG10} with the experimental data for the \SI{3}{\watt} sample.}
	\label{figure4}
\end{figure}

The analysis of the cation distribution and valence is restricted to Ni and Fe since those elements show a significant XMCD signal and thus contribute to the magnetic properties in the NiZAF thin films. In Fig.\,\ref{figure4} the XMCD spectra of the Zn series at the Fe and Ni L$_{3,2}$ edges are shown.\\
At a first glance, the XMCDs at the Fe L$_{3,2}$ edges in Fig.\,\ref{figure4}(a) exhibit a similar shape for the entire series with the exception of the \SI{6}{\watt} sample. Since the L$_{2}$ edge shows no significant differences between the samples besides the intensity, only the L$_{3}$ edge is discussed in detail. The two negative peaks stem from octahedrally coordinated Fe. The peak at lower energies contains contributions from both \Fetwo\, and \Fethree. The one at higher energies solely stems from \Fethree. The positive peak is attributed to tetrahedral \Fethree, which is thus coupled antiparallel to the octahedral Fe, as known for nickel ferrite in general. These peak assignments are based on multiplet ligand field theory \cite{SG10} and were reported in several publications before \cite{L20, K14}. \\
The strongest deviations of the Fe L$_{3}$ edge in the entire series is evident for the \SI{6}{\watt} sample, since an additional feature between the first two peaks arises and its overall shape and relative intensities differ. The differences between the other three sample are more subtle and are mainly seen in the positive tetrahedral \Fethree\, peak, where the intensity is increased for the \SI{5}{\watt} sample. This is in a sense remarkable, because Zn is expected to predominantly occupy tetrahedral sites, however, here an increase in Zn increased Fe on the tetrahedral site. Furthermore, the first negative peak of the XMCD is lowest for the \SI{3}{\watt} sample suggesting the lowest contribution of \FeOhtwo\, (disregarding the \SI{6}{\watt} sample). Also here the increasing \Zntwo\, content counterintuitively leads to an increase of the unwanted \Fetwo. Even though the signal for \FeOhtwo\, seems to be lower in the \SI{6}{\watt} sample, the relative peak height compared to the octahedral \Fethree\, is higher. This correlates well with the results from the magnetic analysis, since the presence of \FeOhtwo\, increases damping by hopping mechanisms and was found to be at a minimum for the \SI{3}{\watt} sample. In turn, these findings indicate that already at the lowest possible Zn concentration the beneficial aspects of the Zn incorporation are reached and a further increase in Zn content counteracts both the suppression of tetrahedral \Fethree\, as well as \Fetwo. \\ 
A comparison of the XMCD spectra at the Ni L$_{3,2}$ edges is shown in Fig.\,\ref{figure4}(b). Analogous to the Fe spectra the overall shape is rather similar for each sample, especially the L$_2$ edge only shows small variations. Focusing on the L$_3$ edge the \SI{6}{\watt} sample is again conspicuously different exhibiting an additional positive peak at the low energy side of the negative peak, which is absent in the other samples. According to multiplet ligand field theory \cite{D11}, this peak is assigned to \NiTdtwo. Since the presence of \NiTdtwo\, is known to increase damping by unquenched orbital momenta, this will further contribute to the increased FMR linewidth in the \SI{6}{\watt} sample. A direct comparison of the other three samples with respect to \NiTdtwo\, is not possible, since the feature is absent. Solely the magnetic contribution from \NiOhtwo, which is responsible for the overall shape of the magnetic dichroism spectrum can be discussed. The octahedral contributions seem to be lesser in the \SI{5}{\watt} sample in comparison to the two samples with lower Zn concentration. Also here, the role of increased Zn content is surprising, since the magnetic signal of Ni on octahedral sites is reduced. \\
An analysis of the cation distribution at the Fe and Ni  L$_{3,2}$ edges shows that increasing the sputter power of the ZnO target up to \SI{5}{\watt} leads to small but noticeable changes in the site occupancy at both edges. In case of Fe an increase in \Fethree\, at tetrahedral sites and in case of Ni a reduction of the \NiOhtwo\, signal are apparent. Since the stoichiometry was not changed drastically by the Zn variation, this is a first indication that Zn also occupies octahedral sites. Increasing the Zn amount even more, makes the changes to the XMCD spectra even more pronounced. Moreover, a sign for \NiTdtwo\, appears, which suggest that Ni moved to tetrahedral sites. These findings are highly unexpected, since according to theory \cite{D11} and previous studies on NiZAF \cite{E17,L20} the addition of Zn should prevent Ni from occupying tetrahedral sites and not promoting it. Obviously, a critical Zn concentration exists in this quaternary oxide, above which the influence of Zn is detrimental to the desired magnetic properties.  \\
So far only direct comparison between the samples was given using multiplet ligand field theory to correlate valence and occupancy with the peaks observed in the XMCD spectra. By applying multiplet ligand field simulation using CTM4XAS \cite{SG10}, a suggestion for the respective percentages of the elemental contributions using a weighted linear combination can be calculated. This has already been done in a previous publication \cite{L20}, where the details of the used parameters are given in detail, please also refer to \cite{K14,PLH11,PMG09,K79}. In Fig.\,\ref{figure4}(c) a comparison of the experimental XMCD spectra and the respective simulation of the \SI{3}{\watt} sample at the Fe L$_{3,2}$ edges is shown. The experimental data is in good agreement with the simulation at the L$_{3}$ edge, which will also be the focus of discussion. The simulation yields the following percentages: \SI{33}{\percent} of \FeOhtwo\, \SI{26}{\percent} of octahedral and \SI{41}{\percent} of tetrahedral \Fethree, respectively. Compared to percentages of \SI{21}{\percent} obtained in Ref.\,\cite{L20} this is an increase of \FeOhtwo\, by approximately \SI{5}{\percent}. Since the amount of octahedral \Fethree\, is equal within error bars, these \SI{5}{\percent} must have shifted from tetrahedral \Fethree\, to \FeOhtwo. \\
Summarizing our findings, the Zn-rich NiZAF \cite{L20} has a decreased Curie temperature below \SI{400}{\kelvin}, while the saturation magnetization is increased. As a comparison of the XMCD spectra showed this is accompanied by a reduction of \FeOhtwo. However, by adapting the Zn content a sample with a slightly higher out of plane lattice parameter of $a_{\perp} = (8.50 \pm 0.01)$\,\si{\angstrom} in comparison to $a_{\perp} = (8.49\pm 0.01)$\,\si{\angstrom} as well as a decrease in coercivity lower than $\mu_0H_{\text{c}} \sim 0.2 $\,\si{\milli\tesla} at \SI{300}{\kelvin} was fabricated. Interestingly, a comparison of the linewidth at \SI{9.5}{\giga\hertz} with \cite{L20} suggests a lower FMR linewidth in case of the sample from the Zn series, even though the XMCD spectrum indicates a higher signal of \FeOhtwo. However a change in FMR linewidth at one fixed frequency does not allow a solid conclusion about the magnetic damping. Therefore, and to identify the different contributions to linewidth broadening, a thorough investigation by broadband FMR is in order, which will be discussed in the next section.
\section*{VI. Broadband FMR: thickness series}

\begin{figure*}[ht]
	\centering
	\includegraphics[width=0.98\textwidth]{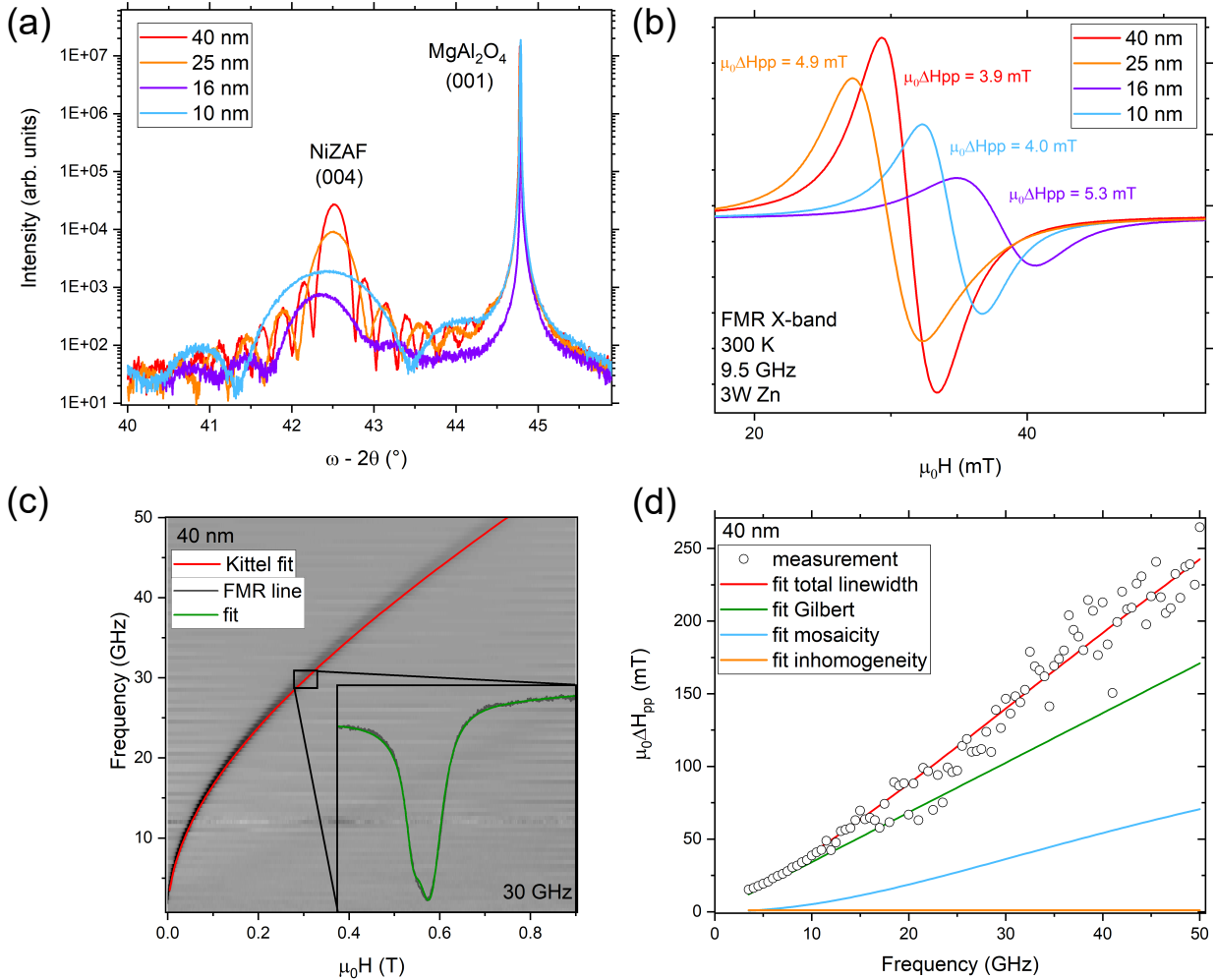}
	\vspace{-0.2cm}
	\caption{In (a) the symmetric $\omega-2 \theta$ scans and in (b) the X-band FMR at \SI{300}{\kelvin} and \SI{9.5}{\giga\hertz}, for each sample from the thickness series are shown. (c) and (d) show a 2D gray-scale plot of the resonance position and the linewidth with broadening contributions, respectively over a frequency range of up to \SI{50}{\giga\hertz} for a \SI{40}{\nano\meter}-thick sample. The inset in (c) depicts the resonance line at \SI{30}{\giga\hertz}.}
	\label{figure5a}
\end{figure*}

\begin{figure*}[ht]
	\centering
	\includegraphics[width=0.98\textwidth]{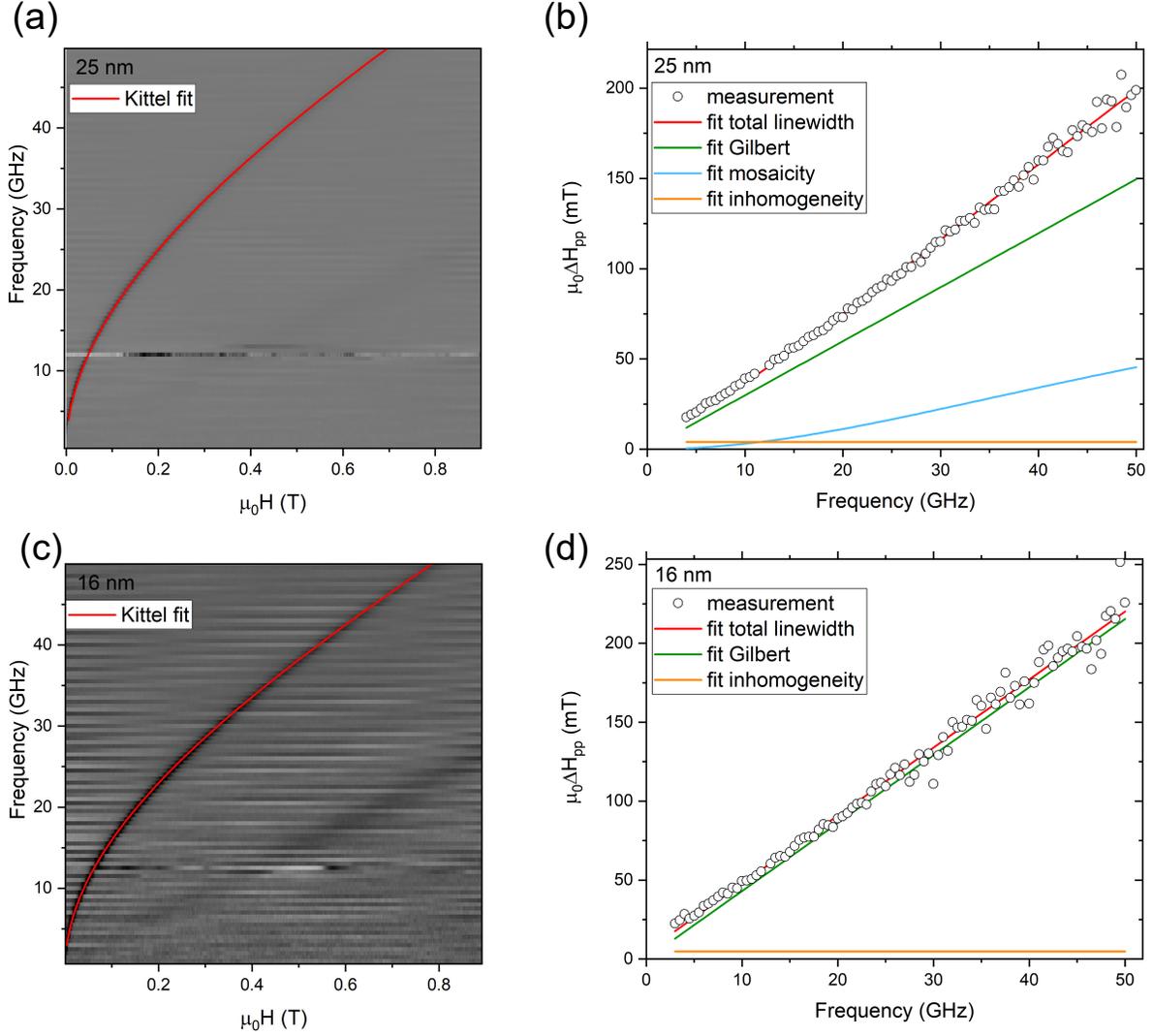}
	\vspace{-0.2cm}
	\caption{In (a) a  2D gray-scale plot of the resonance position and in (b) the linewidth with broadening contributions over a frequency range of up to \SI{50}{\giga\hertz} for a \SI{25}{\nano\meter} thick sample is shown. (c) and (d) depict the results of a \SI{16}{\nano\meter} thick sample, respectively.}
	\label{figure5b}
\end{figure*}

The preliminary analysis with X-band FMR indicates that the lowest damping occurs in the \SI{40}{\nano\meter}-thick \SI{3}{\watt} sample. However, the reported ideal sample thickness for the Zn-deficient NiZAF was \SI{16}{\nano\meter} \cite{E17} and the Zn-rich sample was \SI{25}{\nano\meter} thick \cite{L20}. Therefore, a thickness series has been grown while keeping all other preparation parameters constant to check if the magnetic damping could be lowered further. \\
The results of the XRD measurements of this thickness series are shown in Fig.\,\ref{figure5a}(a). It can be seen, that the structural quality of the samples is retained for a thickness reduction down to \SI{10}{\nano\meter}. Additionally, the diffractograms nicely show the correlation between the sample thickness and the Laue oscillations. The position of the (004) NiZAF peak is identical for the \SI{25}{\nano\meter} and \SI{40}{\nano\meter} thick samples. However, reducing the thickness further shifts the peak to lower angles around \SI{42.43}{\degree} indicating an increased $a_{\perp}$ and thus increased strain. From SQUID magnetometry a small coercivity around $\mu_0H_{\text{c}} \sim 0.2 $\,\si{\milli\tesla} at \SI{300}{\kelvin} and a Curie temperature well above \SI{400}{\kelvin} could be determined for all measured thicknesses (not shown). The saturation magnetization does not show a systematic variation with thickness within experimental uncertainties. The values range from a $M_{\text{s}} = (185 \pm 19)$\,\si{\kilo\ampere/\meter} for the \SI{10}{\nano\meter} sample to a $M_{\text{s}} = (147 \pm 19)$\,\si{\kilo\ampere/\meter} for the \SI{16}{\nano\meter} sample. A comparison of the X-band FMR results at \SI{300}{\kelvin} is provided in Fig.\,\ref{figure5a}(b) demonstrating that the \SI{40}{\nano\meter} sample has the lowest linewidth. However, all other thicknesses show only slightly different results and no conclusive trend with thickness is visible. Therefore, a detailed analysis of the magnetic properties by broadband FMR up to \SI{50}{\giga\hertz} is required to verify and disentangle the various possible contributions to linewidth broadening. \\
A 2D gray-scale plot of the broadband FMR measurements of the \SI{40}{\nano\meter} sample up to \SI{50}{\giga\hertz} is shown in Fig.\,\ref{figure5a}(c). From \SI{30}{\giga\hertz} on the plot becomes visibly blurry. This broadening is caused by a weak second resonance mode overlapping with the main mode. The inset of Fig.\,\ref{figure5a}(c) shows the fitted resonance line at \SI{30}{\giga\hertz}, in which at least a second line is evident. A possible explanation for the appearance of a second or even more resonance lines could be inhomogeneities, i.e. areas with slightly different magnetic properties leading to a different resonance field. A consequence of the appearance of the second line is the strong scatter in the data of the extracted linewidth at higher frequencies. From the frequency-dependent resonance fields of the main peak a g-factor of $g = 2.17 \pm 0.09$ was determined by using the Kittel fit \cite{Z07}. In Fig.\,\ref{figure5a}(d) the linewidth is plotted over the same frequency range with the respective broadening contributions. In addition to the Gilbert damping, which is linear in frequency, another contribution has to be introduced that considers effects caused by mosaicity (due to sample inhomogeneity). Furthermore, a small frequency-independent inhomogeneous damping has to be added to satisfyingly fit the experimental data. Note that, no contribution from two magnon scattering could be observed, which occurred in NiZAF thin films as reported in Ref.\,\cite{L20}. Nonetheless, the obtained Gilbert damping parameter of $\alpha = 9 \times 10^{-3}$ is higher than for the sputtered Zn-rich NiZAF in Ref.\,\cite{L20}. This was unexpected, since the X-band FMR at \SI{9.5}{\giga\hertz} indicated a lower linewidth than previously determined. However, this highlights the necessity of measuring broadband FMR above \SI{30}{\giga\hertz} to be able to accurately determine the Gilbert damping parameter. \\
The same measurements were performed on the \SI{25}{\nano\meter} thick samples, which showed no indication for a second resonance line as can be seen in Fig.\,\ref{figure5b}(a). This sample has a g-factor of $g = 2.15 \pm 0.09$ and the frequency dependence of the linewidth including the different contributions obtained by fitting are shown in Fig.\,\ref{figure5b}(b). The experimental data can only be fitted with the linear Gilbert term plus a mosaicity and  a small inhomogeneous offset. A contribution of two magnon scattering is again absent. The obtained Gilbert damping parameter of $\alpha = 7.8 \times 10^{-3}$ is lower in comparison to the \SI{40}{\nano\meter}-thick sample and also mosaicity and inhomogeneous broadening are reduced. Since \SI{16}{\nano\meter} is the proposed best thickness with the lowest damping in case of NiZAF grown with PLD \cite{E17}, a \SI{16}{\nano\meter} sample was analyzed as well. The results of the 2D gray-scale plot and the fitted linewidth contributions are shown in Figs.\,\ref{figure5b}(c) and (d), respectively. The quantitative analysis yields $g = 2.18 \pm 0.09$. The frequency-dependent linewidth could be fitted without mosaicity, however the linear Gilbert term is increased to $\alpha = 11.4 \times 10^{-3}$ compared to the thicker samples; a small inhomogeneous offset also remains. \\
The broadband FMR showed that by reducing the thickness not only the structural but also the magnetic quality of the samples could be retained confirming the reproduceability of the sputter process with the found optimal preparation conditions. The unsystematic change of the linewidth with thickness is uncharacteristic to previous reports on NiZAF \cite{E17}, where a clear reduction with thickness was found. However, in particular for very thin samples, even a low amount of defects already has a big impact on site occupancy. Additionally, for \SI{16}{\nano\meter} and \SI{10}{\nano\meter} the higher strain inferred from XRD could be another cause for the increased damping. Nonetheless, the thickness series showed that a minimum in the Gilbert damping is found for a thickness of \SI{25}{\nano\meter}. In agreement with Ref.\,\cite{L20}, where RMS grown NiZAF with a similar thickness exhibited a comparable Gilbert damping of $\alpha = 6.8 \times 10^{-3}$. Nevertheless, the cationic concentrations were different, see Tab.\,\ref{table1}. Additionally, it has to be considered that the samples here are strained even more and have a clearly higher Curie temperature up to \SI{450}{\kelvin}, while retaining a very low damping. Furthermore, the even lower g-factor of $g = 2.15 \pm 0.09$ compared to $g = 2.29 \pm 0.09$ \cite{E17} would suggest a reduced amount of \NiTdtwo, since they contribute mostly by unquenched orbital momenta. \\
On the \SI{25}{\nano\meter} sample with lowest damping additional measurements were performed to check the anisotropy in-plane along the azimuthal angle in a range from \SIrange{0}{360}{\degree} and out-of plane along the polar angle in range from \SIrange{-20}{90}{\degree} at a frequency of \SI{15}{\giga\hertz} (not shown). To fit the angular dependences and determine the different contributions to anisotropy, the resonance equation analogous to \cite{L20,F98} was used. 
The azimuthal angular dependence indicates a four fold symmetry as expected from a cubic system resulting in a cubic anisotropy of $2K_{4\parallel}/M_\mathrm{s} = 2.2$\,\si{\milli\tesla},  with an additional small uniaxial contribution of $2K_{\parallel}/M_\mathrm{s} = 0.8$\,\si{\milli\tesla}. These findings are contrary to Refs.\,\cite{E17} and \cite{L20}, which reported higher values of \SI{10}{\milli\tesla} and a slight shift towards a tetragonal crystal symmetry, respectively. The polar angular dependence revealed an out-of-plane anisotropy of \SI{3.5}{\tesla}, which is considerable higher than reported before and can be explained by the increased $c/a$ ratio. It is also in quantitative agreement with the SQUID measurements at \SI{300}{\kelvin} in in-plane and out-of-plane orientation (not shown).\\
Having now optimized the Zn content and thickness, the remaining parameter is the strain of the NiZAF film, which can be controlled by the Al content as discussed before \cite{E17} and in bulk samples \cite{L13}. Therefore, the Al concentration is varied in a next step using the optimized parameters from the Zn series.
 
\section*{VII. Dependence of Al content}

\begin{figure*}[ht]
	\centering
	\includegraphics[width=0.98\textwidth]{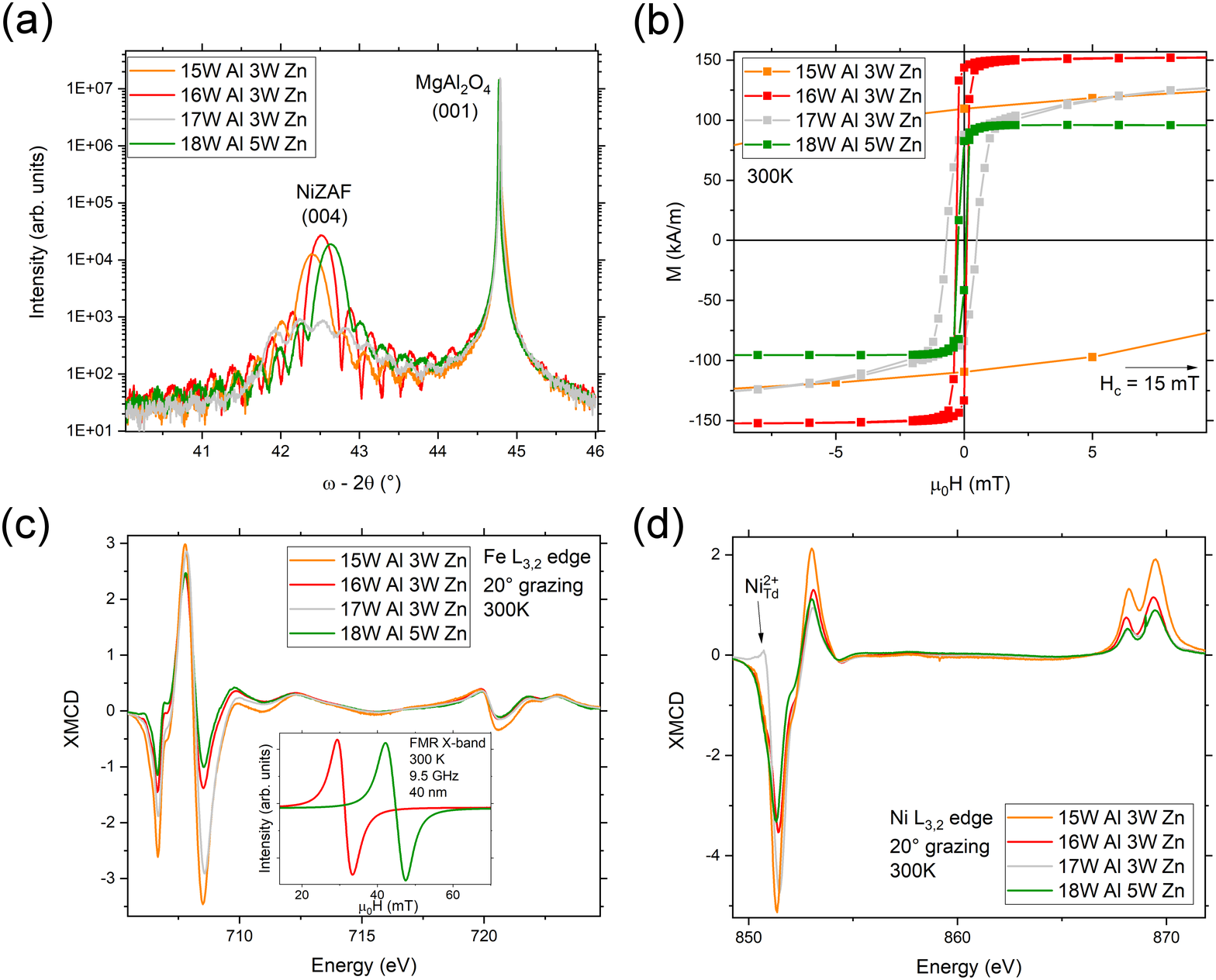}
	\vspace{-0.2cm}
	\caption{Static and magnetic analysis of the Al series. In (a) the symmetric $\omega-2 \theta$ scans and in (b) the $M(H)$ curves at \SI{300}{\kelvin} and \SI{5}{\tesla}, for each sample from the Al series are shown. (c) and (d) depict the XMCD spectra at \SI{300}{\kelvin} in \SI{20}{\degree} grazing at the Fe L$_{3,2}$ edges Ni L$_{3,2}$ edges, respectively for the each sample from the Al series. The inset in (c) shows a comparison of the FRM lines of two samples from the Al series.}
	\label{figure6}
\end{figure*}

The symmetric $\omega-2 \theta$ scans measured by XRD of the four samples from the Al series are shown in Fig.\,\ref{figure6}(a). Three samples have a Zn amount of \SI{3}{\watt} with a varied Al amount between \SIrange{15}{17}{\watt} and a promising sample where both doping parameters were changed is analyzed as well. The strongest changes in structure are apparent in the sample sputtered with \SI{17}{\watt}, where the crystal quality seems most disordered. Even though, Laue oscillations are visible, indicating a smooth surface, the crystal quality is greatly reduced. The other three samples show a dependence of the lattice parameter on the Al content. By increasing the Al content the (004) film peak moves to higher angles, decreasing the lattice parameter and therefore reducing strain. The sample in which both dopants were adapted (green line) has an increased peak position of \SI{42.65}{\degree}, corresponding to $a_{\perp} = (8.47 \pm 0.01)$\,\si{\angstrom}.\\
The $M(H)$ curves at \SI{300}{\kelvin} of the samples from the Al series are shown in Fig.\,\ref{figure6}(b). They show that the best sample from the Zn series still has the lowest $\mu_0H_{\text{c}} = 0.2 $\,\si{\milli\tesla} as well as the highest saturation magnetization of $M_{\text{s}} = (159 \pm 16)$\,\si{\kilo\ampere/\meter} at \SI{300}{\kelvin}. The \SI{15}{\watt} sample shows the biggest deviation, where the coercivity considerably increased up to $\mu_0H_{\text{c}} = 15.0 $\,\si{\milli\tesla} as denoted in the figure. In comparison the most similar result to the best sample of the Zn series (red line) is obtained from the sample where both the Zn and Al content were adapted (green line). In the Al series the $M(T)$ behavior is comparable to the Zn series and the Curie temperature remains well above \SI{400}{\kelvin} (not shown). However, the sample where both doping parameters were changed (green line) retains a good structural quality as well as coercivity, but has a decreased Curie temperature of $T_{\text{C}} = (385 \pm 2)$\,\si{\kelvin} similar to \cite{L20}. Additionally, the saturation magnetization is significantly lowered to $M_{\text{s}} = (110 \pm 16)$\,\si{\kilo\ampere/\meter}. \\
In a next step, changes in the cationic occupation and/or valence are to be researched. XMCD spectra were recorded at the Fe and Ni L$_{3,2}$ edges under the same conditions as for the Zn series. The results of the Fe L$_{3,2}$ edges are shown in Fig.\,\ref{figure6}(c) and the discussion is again focused on the L$_3$ edge. The elemental contributions of Fe are very similar in the best sample from the Zn series as well as the sample in which both doping parameters were changed (red and green line). In the other two samples the amount of octahedral Fe is heavily changed, in particular for \Fethree. XMCD spectra at the Ni L$_{3,2}$ edges shown in Fig.\,\ref{figure6}(d) also reveal significant changes upon variation of the Al content. The two samples with higher structural perfection also have similar XMCD spectra. In comparison the \SI{15}{\watt} and \SI{17}{\watt} sample exhibit an increased \NiOhtwo\, peak. Additionally, the \SI{17}{\watt} sample has a positive peak at lower energies as well, which indicates the presence of a significant amount of \NiTdtwo. The similarity of the XMCD spectra at both edges of the best sample from the Zn series and the sample where both dopants were changed (red and green line) coincides with the structural and magnetic properties, which were significantly better compared to the remaining samples of the Al series. Furthermore, the deviating structural or magnetic properties apparent in the \SI{17}{\watt} and \SI{15}{\watt} sample, respectively can be explained as well. The XMCD spectra at the Fe L$_3$ edge have an increased amount of octahedral Fe, in particular the \SI{15}{\watt} sample. As a result the coercivity was drastically increased. Additionally, the \SI{17}{\watt} sample shows the presence of \NiTdtwo, which leads to strong changes in the XRD and in combination with the higher amount of \FeOhtwo\, to an increased coercivity. The inset in  Fig.\,\ref{figure6}(d), shows a comparison of the FMR lines of the best samples from the Al series. The sample where both doping parameters were changed achieves a similar small linewidth of $\mu_0\Delta H_{\text{pp}} = (4.6 \pm 0.2)$\,\si{\milli\tesla} as obtained for the best samples from the Zn series. However, due to the reduced strain by the higher amount of Al the anisotropy decreased and the resonance moves to higher fields of \SI{44.8}{\milli\tesla}. The \SI{15}{\watt} and \SI{17}{\watt} samples are not discussed further, since the first shows an increased coercivity while the second has a drastically reduced crystal quality in addition to the presence of \NiTdtwo.\\
A summary of the parameters obtained from the analysis of the Al series can be seen in Tab.\,\ref{table2}. In the table the elemental ratios determined by RBS using a \SI{10}{\mega\electronvolt} $^{12}$C$^{3+}$ ion beam are depicted as well. However, a distinct dependence of the ratios on the Al content as seen for the Zn series is not as apparent. The Al\,:\,Fe ratio increases with increasing sputter power of the Al target from $0.58$ to $0.68$. Also the Al\,:\,Ni ratio shows similar results with an increase from $1.06$ to $1.20$. No dependence for the Al\,:\,Zn ratio could be determined. The Ni\,:\,Fe ratio is rather constant around $0.56$, similar to the Zn series. \\
The analysis of the Al series experimentally confirmed, that the strain in the material can be controlled by adjusting the Al amount, as predicted from theory \cite{D11}. However, a deviation of the Al content from the optimized parameters of the Zn series resulted either in the loss of structural quality or increased coercivity as well as the presence of \NiTdtwo. More importantly, a comparison of the best sample from the Zn series with a sample where both Al and Zn were varied showed a decrease in saturation magnetization and Curie temperature. However, the release in strain lead to a decreased anisotropy while retaining a similar small linewidth. The comparison suggests that due to the complex interplay of the elements in NiZAF several favorable combinations in the parameter space of the growth conditions can be found to achieve ferromagnetic insulators with low magnetic damping for spintronic applications.

\begin{table*}[ht]
	\centering
	\caption{Summary of the structural and magnetic properties of the samples from the Al series. The respective measuring errors are given in the first row.}
	\begin{tabular}{c|cccc|c|cc|ccc}
		\hline \hline
		Growth & \multicolumn{4}{c|}{RBS} & XRD & \multicolumn{2}{c|}{SQUID} & \multicolumn{2}{c}{FMR} \\ 
		Zn {(}\si{\watt}{)} &Al:Fe & Al:Zn& Al:Ni &Ni:Fe   & (004) reflex {(}\si{\degree}{)} & $M_{\text{s}}${(}\si{\kilo\ampere/\meter}{)} & $\mu_0H_{\text{c}}${(}\si{\milli\tesla}{)} & $\mu_0 H_{\text{res}}${(}\si{\milli\tesla}{)} & $\mu_0\Delta H_{\text{pp}}${(}\si{\milli\tesla}{)} \\ \hline \hline
		$\pm 0.5$ {(}\si{\watt}{)} & & &  &  & $\pm 0.02$ {(}\si{\degree}{)} & $\pm 10${(}\si{\percent}{)} & $\pm 0.2${(}\si{\milli\tesla}{)} & $\pm 0.2${(}\si{\milli\tesla}{)} & $\pm 0.2${(}\si{\milli\tesla}{)} \\
		15.0          &0.58&2.28&1.06& 0.55     & 42.42     &  190    &   0.8      &         &          \\ 
		16.0          &0.62&2.69&1.06&  0.59     & 42.50      & 159      & 0.2     & 31.3   & 3.9            \\ 
		17.0          &0.63&2.34&1.12&  0.56   & 42.41     &  142     &   15.0      &        &            \\ 
		18.0 (5W Zn)  &0.68&2.13&1.20&  0.56    &  42.65     &  110    &   0.2     &  44.8     & 4.6        \\ 
	\end{tabular}
	\label{table2}
\end{table*}

\section*{VI. Conclusion}
In this work the systematic variation of the Zn and Al concentrations in Zn/Al doped nickel ferrite lead to new insights regarding the interplay of cation distribution, site occupancy, and magnetic damping. Three different sample series were investigated to identify the effects of Zn and Al content as well as thickness, due to its reported correlation to magnetic damping \cite{E17}.\\
From the Zn series NiZAF thin films were found to maintain the structural quality over a change in sputter power from \SIrange{3}{6}{\watt} of the ZnO target. The change in Zn content was confirmed by measuring the elemental ratios with RBS. The magnetic analysis with SQUID magnetometry showed a reduction of the saturation magnetization with increasing Zn, while maintaining a small coercivity. Interestingly, the Curie temperature was found to be very robust towards changes in the Zn content and above \SI{400}{\kelvin} throughout the whole series. By increasing Zn the anisotropy decreased, evidenced by the measured higher resonance position and a broadening of the linewidth occurred in FMR. An investigation of the XMCD spectra at the Fe and Ni L$_{3,2}$ edges showed an increase in the tetrahedral \Fethree\, and the presence of \NiTdtwo\, at higher Zn concentrations. Since the stoichiometry did not change drastically, this suggests the presence of octahedral \Zntwo, which is contrary to theoretical studies reporting that Zn solely favors tetrahedral coordination \cite{D11}.To confirm these suspicions XANES on the Zn K edge in combination with simulations would be required, since measurements on the Zn L$_{3,2}$ edges did not allow to draw any meaningful and direct conclusions. Furthermore, a variation of the thickness under the optimized preparation conditions from the Zn series, highlighted the reproduceability of the thin films. The samples from the thickness series had equally promising structural and magnetic properties suitable for a ferromagnetic insulator with a low magnetic damping. \\
A comparison of the results from the structural and magnetic analysis of the Al series showed that the best sample of the Zn series could not be further improved by varying the Al concentration. Only a small variation of the Al amount already lead to drastic changes to either the crystal quality or the magnetic properties. From the XMCD spectra it could be concluded, that only changing both the Zn and Al amount keeps the site occupancy of the Fe and Ni cations relatively unaffected. However, if only one of the parameters is changed the occupancy of Ni and Fe is changed significantly manifesting in either an increased coercivity or loss in structural quality.\\
The best insulating ferromagnet with a low magnetic damping was obtained for a thickness of \SI{25}{\nano\meter} at lowest possible Zn content. A Gilbert damping of  $\alpha = 7.8 \times 10^{-3}$ and a g-factor of $g = 2.15 \pm 0.09$, measured over a frequency range of up to \SI{50}{\giga\hertz}, was achieved. \\
The investigation of the new series in comparison to previous publications \cite{E17,L20} augments previous findings of the site occupancy being the main driving force in determining the magnetic damping in NiZAF thin films. Furthermore, the comparison showed that, there exists more than one stoichiometry for the fabrication of a ferromagnetic insulator with a low magnetic damping suitable for spintronic applications. From this publication a Zn-deficient NiZAF with the so far highest reported strain yielded a Gilbert damping as low as $\alpha = 7.8 \times 10^{-3}$. Despite the heavy strain the Curie temperature was found to be as high as \SI{450}{\kelvin}, unprecedented in sputtered NiZAF thin films. 

\section*{Acknowledgement}
The authors gratefully acknowledge funding by FWF project ORD-49. The X-ray absorption measurements were performed on the EPFL/PSI X-Treme beamline at the Swiss Light Source, Paul Scherrer Institut, Villigen, Switzerland. In addition, support by VR-RFI (Contract No. 2017-00646 9 \& 2019-00191) and the Swedish Foundation for Strategic Research (SSF, Contract No. RIF14-0053) supporting accelerator operation at Uppsala University is gratefully acknowledged. The research leading to this result has been supported by the RADIATE project under the Grant Agreement 824096 from the EU Research and Innovation programme HORIZON 2020.

\end{document}